\documentclass[aps,prl,twocolumn,superscriptaddress,preprintnumbers]{revtex4-2}
\usepackage{braket}
\usepackage{graphicx}
\usepackage{amsmath,verbatim,latexsym,amssymb,indentfirst,mathrsfs,mathtools,amsthm,bbm,bm,hyperref,url}

\newcommand{\II}{\mathcal{I}}
\newcommand{\A}{A}
\newcommand{\F}{F}

\newcommand{\sz}{\ensuremath{\sigma_z}~}

\newcommand{\Tr}{{\rm Tr}}  
 \def\id{\mathbbm{1}}   % Identity
\newcommand{\weak}{{\rm wk}}  
\newcommand{\wv}{{\rm wv}}
\newcommand*{\ketbra}[2]{\lvert #1 \rangle\!\langle #2 \rvert}
\newcommand*{\brakett}[2]{\langle #1 \lvert #2 \rangle}
\newcommand*{\expval}[1]{\left\langle  #1  \right\rangle}
\newcommand{\Det}{{\rm D}}
\newcommand{\nbar}{\bar{n}}

\usepackage{color}
\usepackage[dvipsnames]{xcolor}
\usepackage[normalem]{ulem}
\newcommand{\nicole}[1]{{\color{black}#1}}

\newcommand{\jonathan}[1]{{\color{black}#1}}

\usepackage[title,titletoc]{appendix}

\renewcommand{\refname}{References}

\renewcommand{\thesection}{\Roman{section}}
\renewcommand{\thesubsection}{\Roman{section} \Alph{subsection}}
\renewcommand{\thesubsubsection}{\Roman{section} \Alph{subsection} \arabic{subsubsection}}
\makeatletter
\def\p@subsection{}
\makeatother
\makeatletter
\def\p@subsubsection{}
\makeatother

\usepackage{nameref}
\usepackage{zref-xr}
\zxrsetup{toltxlabel}
\zexternaldocument*{sm}

\newcommand{\povmI}{{\rm I}}
\newcommand{\povmII}{{\rm II}}
\newcommand{\Op}{\mathcal{O}}

\renewcommand\th{ {\rm th} }
\newcommand{\Max}{ {\rm max} }   

\begin{document}
	\title{Weak Measurement of Superconducting Qubit Reconciles Incompatible Operators}
	
	
	\author{Jonathan T. Monroe}
	\email[]{j.monroe@wustl.edu}
	%\thanks{}
	\affiliation{Department of Physics, Washington University, St. Louis, MO 63130, USA}
	
	\author{Nicole~Yunger~Halpern}
	\email{nicoleyh@g.harvard.edu}
	\affiliation{ITAMP, Harvard-Smithsonian Center for Astrophysics, Cambridge, MA 02138, USA}
	\affiliation{Department of Physics, Harvard University, Cambridge, MA 02138, USA}
	\affiliation{Research Laboratory of Electronics, Massachusetts Institute of Technology, Cambridge, MA 02139, USA}
	\affiliation{Center for Theoretical Physics, Massachusetts Institute of Technology, Cambridge, MA 02139, USA}
	\affiliation{Joint Center for Quantum Information and Computer Science, NIST and University of Maryland, College Park, MD 20742, USA}
\affiliation{Institute for Physical Science and Technology, University of Maryland, College Park, MD 20742, USA}
	\affiliation{Institute for Quantum Information and Matter, California Institute of Technology, Pasadena, CA 91125, USA}
	
	\author{Taeho Lee}
	\affiliation{Department of Physics, Washington University, St. Louis, MO 63130, USA}
	
	\author{Kater W. Murch}
	\email{murch@physics.wustl.edu}
	%\thanks{}
	\affiliation{Department of Physics, Washington University, St. Louis, MO 63130, USA}

	\begin{abstract}
Traditional uncertainty relations dictate a minimal amount of noise in incompatible projective quantum measurements. However, not all measurements are projective. Weak measurements are minimally invasive methods for obtaining partial state information without projection. Recently, weak measurements were shown to obey an uncertainty relation cast in terms of entropies. We experimentally test this entropic uncertainty relation with strong and weak measurements of a superconducting transmon qubit. A weak measurement, we find, can reconcile two strong measurements' incompatibility, via backaction on the state. Mathematically, a weak value---a preselected and postselected expectation value---lowers the uncertainty bound. Hence we provide experimental support for the physical interpretation of  the weak value as a determinant of a weak measurement's ability to reconcile incompatible operations.
	\end{abstract}
	\preprint{MIT-CTP/5278}
	
	\date{\today}
	\maketitle

Quantum measurements suffer from noise
	that limits precision metrology \cite{Aasi2013,Degen2017}, amplification \cite{Caves1982, Clerk2010}, and measurement-based feedback.
	%References from Ho 2018 which reference precision metrology
	%17: Liebfried 2004: Heisenberg-limited spectroscopy	https://doi.org/10.1126/science.1097576
	%18: Giovannetti 2004: beating standard quantum limit	https://science.sciencemag.org/content/306/5700/1330
	%19: Degen 2017: Quantum sensing			https://journals.aps.org/rmp/abstract/10.1103/RevModPhys.89.035002
	%20: Choi 2018: Sensing with condensed matter		https://arxiv.org/abs/1801.00042
	The minimal amount of noise achievable is lower-bounded in uncertainty relations.
	They highlight how quantum noise arises from 
	disagreement between, or incompatibility of, quantum operations.
	Robertson proved~\cite{Robertson1929}
	the most familiar uncertainty relation:
	the measurement statistics of two observables, $A$ and $B$, must have sufficiently large standard deviations, $\Delta A$ and $\Delta B$, to obey
	\begin{equation}
	\Delta A \; \Delta B \geq \frac{1}{2}|\braket{[A,B]}|.
	\label{robertson_UR}
	\end{equation} 
	Operator pairs with nonzero uncertainty bounds are said to disagree, or to be incompatible. Uncertainty relations quantify the incompatibility. 
	% We show that a weak measurement can reconcile operator disagreement, in a superconducting-qubit experiment. 
	
	Inequality~\eqref{robertson_UR} suffers from shortcomings~\cite{Deutsch_83_Uncertainty}.
	For example, the right-hand side (RHS) depends on a state, through an expectation value. Varying the state appears to vary the disagreement between $A$ and $B$. But the amount of disagreement should depend only on the operators. This objection and others led to the development of \emph{entropic uncertainty relations}
	in quantum information theory~\cite{Coles2017}.
	The variances in Ineq.~\eqref{robertson_UR} give way to entropies, which quantify the optimal efficiencies with which information-processing tasks can be performed \cite{qcqi}. 
	
	An exemplary entropic uncertainty relation was proved in~\cite{Maassen1988}.
	Consider preparing a state $\rho$ and measuring the observable $A$.
	Let $p_a$ denote the probability of obtaining the eigenvalue $a$.
	The probability distribution $\{p_a\}$ has a Shannon entropy  
	$H(A)_\rho \coloneqq -\Sigma_a p_a \log_2 p_a$ 
	equal to the detector's von Neumann entropy. If $H(B)_\rho$ is defined analogously, % the entropies' sum obeys 
	\begin{equation}
	H(A)_\rho + H(B)_\rho > -\log c.
	\label{maassenUffink}
	\end{equation}
$c$ denotes the \emph{maximum overlap} between any eigenstates, $\ket{a}$ and $\ket{b}$, of the observables' eigenstates: 
	$c  \coloneqq  {\rm max}_{a, b}  \left\{|\braket{b| a}|^2\right\}$.
	Inequality~\eqref{maassenUffink} holds for every state $\rho$
	and eliminates state dependence from the bound (RHS), as desired.
	
	The uncertainty relations~\eqref{robertson_UR} and \eqref{maassenUffink} concern only projective, or strong, measurements of observables. 
	\emph{Weak measurements} \cite{Jacobs2006} operate at various measurement strengths. They
have been explored recently in quantum optics~\cite{Kocsis_11_Observing}, cavity quantum electrodynamics (QED)~\cite{Guerlin2007}, and circuit QED~\cite{Vijay2011,hatr13,Murch2013,groen13}. 
	During a weak measurement, the system of interest
	is coupled weakly to a detector, which is then projected~\cite{vonNeumann1932}. 
		% J. von Neumann, Mathematische Grundlagen der Quantemechanik (Springer-Verlag, Berlin, 1932) 
	%ISBN: 9780691178561
	The outcome provides partial information about the system of interest, 
	without projecting the system.
	Weak measurements illuminate quantum dynamics, as in 
	the tracking of the progress of spontaneous emission \cite{Naghiloo2015, Campagne-Ibarcq2016b}, the catching and reversing of quantum jumps \cite{Minev2018}, and observations of noncommuting observables' dynamics \cite{Hacohen-Gourgy2016}.
	
	An entropic uncertainty relation that governs weak measurements
	was proved recently~\cite{Halpern2018}.
	The relation quantifies the disagreement between 
	a strong measurement and 
	the composition of a weak measurement and 
	another strong measurement. 
	We show that the weak measurement can, backacting on the state, 
	reconcile the disagreement between the strong measurements.
	The measurements are performed in a circuit-QED architecture,
	with a superconducting transmon qubit.
	
	Our results reveal a physical significance of weak values.
	A \emph{weak value} is a preselected and postselected expectation value.
	Let $\II = \sum_i \lambda_i  \ketbra{i}{i}$ and 
	$\F  =  \sum_f  f  \ketbra{f}{f}$ denote observables. We assume, throughout this paper, that the eigenspaces are nondegenerate, as we will focus on a qubit. But this formalism, and the theory we test~\cite{Halpern2018}, extend to degeneracies. 
	Consider measuring $\II$, obtaining outcome $\lambda_i$, measuring $F$, and obtaining outcome $f$. 
	Let $A$ denote an observable that commutes with neither $\II$ nor $F$. 
	Which value can be retrodictively ascribed most reasonably to $A$, 
	given the preselection on $\lambda_i$ and the postselection on $f$?
	The weak value~\cite{Aharonov1988}
	\begin{align}
	\label{eq_Weak_Val_Def}
	A_\mathrm{wv} \coloneqq \frac{ \bra{f} A \ket{i} }{ \brakett{f}{i} }  \, .
	\end{align}
	$A_\mathrm{wv}$ can assume anomalous values, which lie outside the operator's spectrum. 
	Weak values’ significance and utility have been debated across theory and experiment~\cite{Leggett1989,Aharonov1990,Dressel2014,Hosten2008, Pang14, Jordan2014}.
	We demonstrate a new physical meaning of the weak value:  
	As a contribution to the uncertainty bound 
	for weak and strong measurements~\cite{Halpern2018},
	$A_\mathrm{wv}$ controls how much weak measurements reconcile incompatibility.
	
	This paper reports on an experimental test of 
	the entropic uncertainty relation for weak and strong measurements~\cite{Halpern2018}.
	We first introduce the experimental platform
	and the dispersive measurements performed in circuit QED. 
	We begin by quantifying two projective measurements' incompatibility with entropies.
	Turning one measurement into a composition---a
	weak measurement followed by a projective measurement---raises
	the overall measurement's entropy by increasing the number of possible outcomes.
	But, under a natural normalization scheme, the weak measurement reduces 
	the sum of the two operations' entropies.
	The entropy sum was bounded % with a weak value 
	in~\cite{Halpern2018},
	whose theory we review and then test experimentally.
	We quantify how the weak measurement backacts on the state.
	Through the backaction, the weak value can lower the uncertainty bound,
	allowing the measurements to agree more. 
	\nicole{In bridging entropic uncertainty relations with weak measurements and superconducting qubits, this work unites several subfields of quantum-information physics, which can benefit from the synergy introduced here. For example, in addition to the fundamental contribution to weak values mentioned above, this work paves the path toward detecting quantum chaos with weak measurements experimentally, being the first experiment to sprout from the considerable theoretical work on leveraging weak measurements to identify chaos~\cite{NYH_17_Jarzynski,NYH_18_Quasiprobability,Swingle_18_Resilience,GonzalezAlonso_19_Out,Dressel_18_Strengthening,Halpern2018,Mohseninia_19_Optimizing,DRMAS_19_Quantum, Arvidsson-Shukur2020}.
}

	%%%%%%%%%%%%%%%%%%%%%%%%%%%%%%%%%%%%
	%%%%%%% EXPERIMENT CONTEXT  %%%%%%%%
	%%%%%%%%%%%%%%%%%%%%%%%%%%%%%%%%%%%%
	
\begin{figure}
		\centering
		\includegraphics[width=0.5\textwidth]{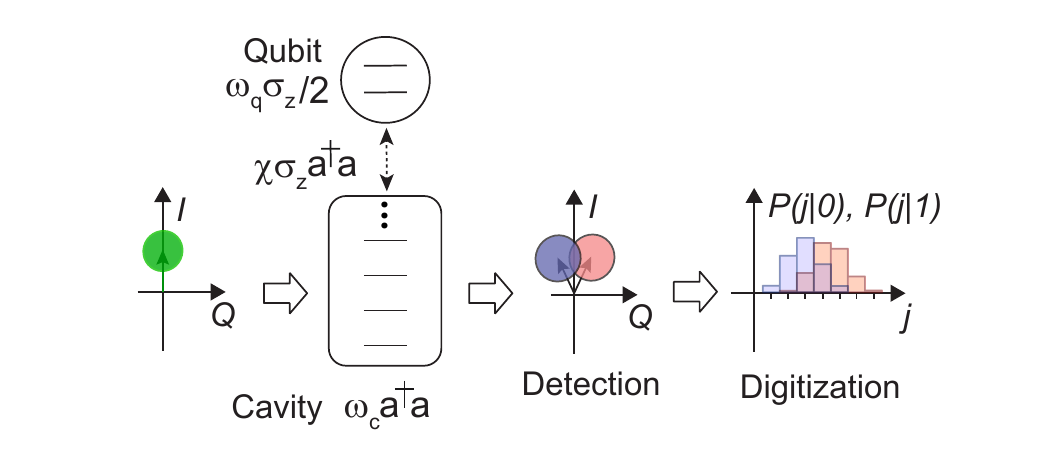}
		\caption{Our experimental setup involves a superconducting transmon qubit coupled dispersively to a microwave cavity. The cavity's state is sketched in phase space, defined by quadratures $I$ and $Q$.
			Coherent states probe the cavity, acquiring a phase shift (red and blue circles) dependent on the qubit's state. The transmitted-probe quadrature that contains qubit-state information is demodulated and digitized into discrete measurement outcomes $j$.
		}
		\label{fig1}
	\end{figure}

	\emph{Experimental context.---}We measure the entropic uncertainty relation with a transmon superconducting qubit. The qubit couples to one mode of the electromagnetic field in a three-dimensional microwave cavity (Fig.~\ref{fig1}). The qubit frequency, $\omega_q/(2\pi) =3.889$ GHz, is far detuned from the cavity frequency, $\omega_c/ (2\pi) =5.635$ GHz, enabling a dispersive interaction. 
	Dispersive interactions do not exchange energy, allowing for quantum-nondemolition measurements.
	The Jaynes-Cummings Hamiltonian in the dispersive limit,
	\begin{equation}
	\label{eq_H_JC}
	H_{\mathrm{JC}}/\hbar = \omega_c a^\dagger a + \frac{1}{2}\omega_q \sigma_z 
	+ \chi a^\dagger a \sigma_z,
	\end{equation}
	governs the measurement dynamics. 
	$a^\dagger$ ($a$) denotes the cavity mode's creation (annihilation) operator, and \sz denotes the Pauli $z$ operator. 
	The final term, $\chi a^\dagger a \sigma_z$, represents the interaction. It effectively changes $\omega_c$ by an amount $\pm\chi= \mp 2\pi (1.5~ \mathrm{MHz})$  dependent on the qubit's state.

	We prepare the cavity probe in a coherent state,
	whose phase shifts in accordance with the qubit's state.
	We perform a homodyne measurement of the field's $Q$ quadrature,
	using a Josephson parametric amplifier.
	The probe state is continuous-variable.
	However, we discretize the possible measurement outcomes into bins
	labeled by $j$.
	
	Outcome $j$ occurs with a probability calculated with
	a positive operator-valued measure (POVM).
	POVMs represent general (not-necessarily-projective) measurements mathematically~\cite{qcqi}.
	A POVM is a set of positive operators  $K_j^\dag K_j > 0$
	that obey the normalization condition $\sum_j K_j^\dag K_j = I$.
	The \emph{Kraus operator} $K_j$ evolves the system-of-interest state:
	$\rho \mapsto K_j \rho K_j^\dag / \Tr ( \rho K_j^\dag K_j )$.
	The denominator equals the measurement's probability of yielding $j$.
	
	Our setup measures the qubit observable $A = \sigma_z$,
	due to Eq.~\eqref{eq_H_JC} and the measurement's homodyne nature~\cite{Boissonneault_09_Dispersive}.
	We can effectively measure other observables $A$
	by rapidly rotating the qubit before and after the interaction.
	\nicole{
	Our phase-sensitive homodyne scheme projectively measures the cavity field along a specific quadrature~\cite{Hatridge_13_Quantum}.  
	If the cavity measurement yields outcome $j$, the qubit state evolves} under the Kraus operator~\cite{Jacobs2006}
	\begin{equation}
	\label{eq_Kraus1}
	K_j  
	=  \left(  \frac{\delta t}{{2 \pi \tau}}  \right)^{1/4} 
	\exp \left(  -\frac{\delta t}{4\tau} [j I - A]^2  \right) .
	\end{equation}
	$\tau$ denotes the characteristic measurement time \cite{webe14}, and the integration time $\delta t$ determines the measurement strength $\delta t/\tau$.
	It depends on system parameters including the mean number of photons in the cavity.
	The Kraus operator's backaction on the qubit state
	will enable the weak measurement to reconcile incompatible operators.

	\begin{figure}
		\centering
		\includegraphics[width=0.5\textwidth]{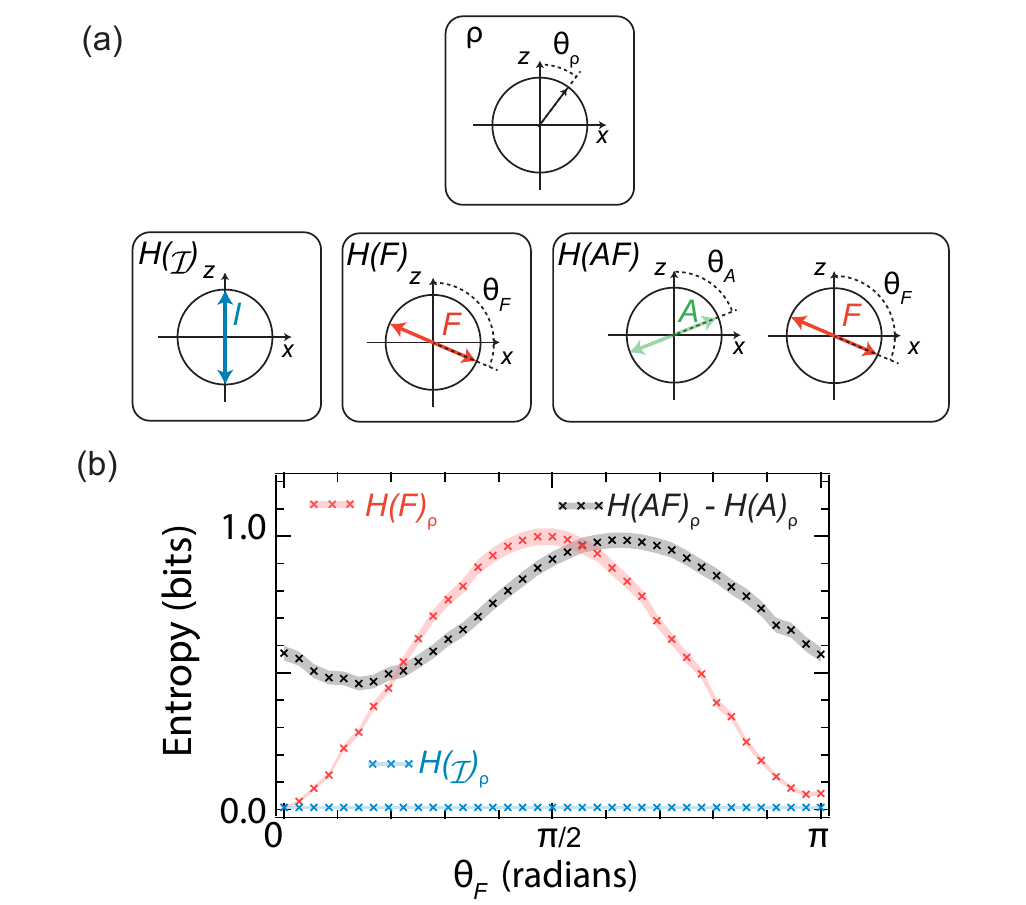}
		\caption{Characterization of entropic uncertainties: 
			(a) We subject a state $\rho$ to one of three measurements. The measurements' entropies are defined as the detectors' von Neumann entropies. 
			(b) Entropies measured for the state $\rho = \ketbra{0}{0}$. \nicole{Bands indicate statistical error from finite sampling (approximately 10,000 repetitions per angle)}. $H(\mathcal{I})_\rho$ and $H(F)_\rho$ characterize projective measurements. $H(AF)_\rho-H(A)_\rho$ quantifies the change, caused by the weak measurement, in the second measurement's entropy, when $\theta_A=\pi/4$. $H(\mathcal{I})_\rho + H(F)_\rho$ maximizes when $\theta_F = \pi/2$, such that $F = X$, while $\II = Z$.  The second measurement's entropy change, $H(AF)_\rho-H(A)_\rho$, maximizes at $\theta_F=0.53\pi$.}   
		\label{fig2}
	\end{figure}

	%%%%%%%%%%%%%%%%%%%%%%%%%%%%%%%%%%%%%%
	%%%%%%%%%%%%%  ENTROPIES   %%%%%%%%%%%
	%%%%%%%%%%%%%%%%%%%%%%%%%%%%%%%%%%%%%%
	\emph{Entropic uncertainties.---}To build intuition, we show how entropic uncertainties arise in our experiment and are modified by weak measurements. 
	First, we define observables $\mathcal{I}$, $F$, and $A$.
	Without loss of generality, we set 
	$\mathcal{I} = \sigma_z = \ketbra{0}{0} - \ketbra{1}{1}$.
	$F$ is represented on the Bloch sphere by
	the axis that lies an angle $\theta_F$ below the $z$-axis,
	at the azimuthal angle $\phi = 0$ [Fig.~\ref{fig2}(a)].
	$A$ is defined analogously, in terms of $\theta_A$.
	
	Consider preparing a state $\rho$ and implementing
	one of the three measurements shown in Fig.~\ref{fig2}(a):
	(i) a projective measurement of $\mathcal{I}$,
	(ii) a projective measurement of $F$, or
	(iii) the composition of a weak $A$ measurement
	and a subsequent projective $F$ measurement.
	We implement a projective measurement experimentally
	by integrating the measurement signal 
	for a time $\delta t \gg \tau$.
	Choosing $\delta t =$ 350 ns and $\tau =$ 6 ns
	realizes a projective measurement with ground-state-fidelity 99\% and excited-state-fidelity 91\%.
	\jonathan{The measurement time, 350 ns, is shorter than the decoherence timescales,   $T_1=50~\mathrm{\mu}$s and $T^*_2=10~ \mathrm{\mu}$s.}
	
	The entropies $H(I)_\rho$ and $H(F)_\rho$ are defined as follows.
	In each of many trials, we prepare a qubit state $\rho$ and measure $\II$. 
	From the outcome statistics, we infer the probabilities
	$p_i = \bra{i} \rho \ket{i}$.
	From $\{ p_i \}$, we calculate $H(\II)_\rho$.
	We determine $H(F)_\rho$ analogously. 
	For the data shown in Fig.~\ref{fig2}(b),
	$\rho = \ketbra{0}{0}$.
	The entropies' sum peaks at $\theta_F = \pi/2$,
	signaling the maximal incompatibility of $\sigma_z$ with $\pm \sigma_x$.
	$\II$ and $F$ coincide at $\theta_F = 0$, where the entropy sum minimizes.
	The sum $\gtrsim 0$ because the measurements have finite fidelities.
	
	Figure \ref{fig2}(b) displays also the entropy of 
	the joint $AF$ measurement, for $\theta_A = \pi/4$. 
	In each of many trials, we prepare $\rho$, measure $A$ weakly,
	and measure $F$ projectively. 
	From the frequencies of the outcome tuples $(j,f)$, we infer the joint probabilities $p_{j,f} = \bra{f} K_j \rho K_j^\dagger \ket{f}$.
	On the distribution, we calculate the entropy $H(AF)_\rho$.
	
	$j$ assumes one of $\approx 2^4$ possible values,
	so the weak measurement raises the entropy by $\approx 4$ bits.
	Aside from this increase, measuring $A$ \emph{reduces} entropy sum
	when $\theta_F = \pi / 2$, where
	$F = \sigma_x$ disagrees maximally with $\mathcal{I} = \sigma_z$. 
	To highlight this effect, we normalize $H(AF)_\rho$, displaying the difference $H(AF)_\rho-H(A)_\rho$ in Fig.~\ref{fig2}(b). 
	The weak $A$ measurement reconciles the two operators,
	as we now quantify in more detail.

	%%%%%%%%%%%%%%%%%%%%%%%%%%%%%%%%%%%%
	%%%%%%%%%%%% THEORY  %%%%%%%%%%%%%%%
	%%%%%%%%%%%%%%%%%%%%%%%%%%%%%%%%%%%%
	
	\emph{Theory.---}We briefly review the derivation of the entropic uncertainty relation for weak and strong measurements~\cite{Halpern2018}.
	For convenience, we reuse the definitions in the previous two sections.
	The theory generalizes, however, beyond circuit QED and qubits.
	Recall two of our POVMs,
	(i) a projective $\II$ measurement and
	(iii) the composition of a weak $A$ measurement and
	a projective $F$ measurement.

	We formalize a general weak measurement as follows.
	A detector is prepared in a state $\ket{\Det}$,
	coupled to the system's $A$ weakly via a unitary $V$, 
	and measured projectively.
	If outcome $j$ obtains, the system evolves under the Kraus operator
	$K_j  =  ( {\rm phase} ) \bra{j} V \ket{\Det}$.
	Taylor-approximating in the coupling strength 
	yields~\footnote{
		% < f >
		Our $g_j$ is defined as the $g_j / \sqrt{p_j}$ in~\cite{Halpern2018}.}
	% < /f >
	\begin{align}
	\label{eq_Kraus2}
	K_j  = \sqrt{ p_j } \left\{ 
	I  +  g_j A   +  O \left( [ g_j ]^2 \right)  \right\} .
	\end{align}
	% The term %% NYH: First, these words contribute only to the word count. Second, p_j isn't a term.
	$p_j$ equals the probability that, 
	if the detector is prepared in $\ket{\Det}$ and does not couple,
	the measurement yields $j$.
	$g_j$ quantifies the interaction strength and is defined, 
	via the Kraus operators' unitary invariance~\cite{qcqi}, to be real.
	Comparing with Eq.~\eqref{eq_Kraus1}, we calculate 
	the cavity-QED $p_j$ and $g_j$ in the Supplemental Material~\cite{Suppl_Mat}.
	
	The entropic uncertainty relation for weak and strong measurements 
	is proved as follows.
	We begin with a generalization, to POVMs, of 
	the entropic uncertainty relation~\eqref{maassenUffink}~\cite{Tomamichel_12_Framework,Krishna_01_Entropic}.
	POVMs (i) and (iii) are substituted into the relation.
	The left-hand side (LHS), 
	$H( \II )_\rho  +  H (AF)_\rho$, consists of entropies
	defined as in the previous section.
	The entropies quantify the average uncertainties about
	the POVMs' outcomes.
	
	The uncertainty relation's RHS contains a maximum overlap, 
	similarly to Ineq.~\eqref{maassenUffink}.
	This overlap, however, is between POVM elements.
	In its raw form, the RHS cannot be straightforwardly inferred from experiments.
	Therefore, the bound was Taylor-approximated in the weak coupling, 
	$g_j \sqrt{p_j}$.
	The entropic uncertainty relation for strong and weak measurements results:
	\begin{align}
	\label{EUR}
	%LHS
	& H(\II)_\rho + H(\A\F)_\rho \geq 
	\\ \nonumber &
	%RHS
	\mathrm{min}_{i,j,f} \left\{ 
	-\log_2  (  p_{f|i}  \:  p_j  )
	- \frac{2}{ \ln 2 }  \:
	\Re(g_j A_\mathrm{wv}) 
	+  O  \left(  p_j [ g_j ]^2  \right)
	\right\}.
	\end{align}
	
	The bound contains two non-negligible terms. 
	The zeroth-order term depends on the eigenstate overlap 
	$p_{f | i} = | \brakett{f}{i} |^2$
	in the entropic uncertainty relation~\eqref{maassenUffink} for projective measurements.
	\jonathan{The first-order-term depends on the weak value's real part, $\Re (A_\mathrm{wv})$
	[Eq.~\eqref{eq_Weak_Val_Def}].}
	Positive weak values tend to achieve the minimum, we find,
	leading to a negative $A_\mathrm{wv}$ term.
	\jonathan{The negative sign comes from the negative sign in (the generalization, to POVMs, of) Ineq. (\ref{maassenUffink})}.
	The term lowers the bound, enabling the POVMs to agree more,
	as our experiment shows.
	
	%%%%%%%%%%%%%%%%%%%%%%%%%%%%%%%%%%%%
	%%%%%%%%%%%%%%%%%% METHODS  %%%%%%%%%%%
	%%%%%%%%%%%%%%%%%%%%%%%%%%%%%%%%%%%%

	%%%%%%%%%%%%%%%%%%%%%%%%%%%%%%%%%%%%
	%%%%%%%%%%%%%%%%%% RESULTS %%%%%%%%%%%%
	%%%%%%%%%%%%%%%%%%%%%%%%%%%%%%%%%%%%

	\begin{figure}
		\centering
		\includegraphics[width=0.5\textwidth]{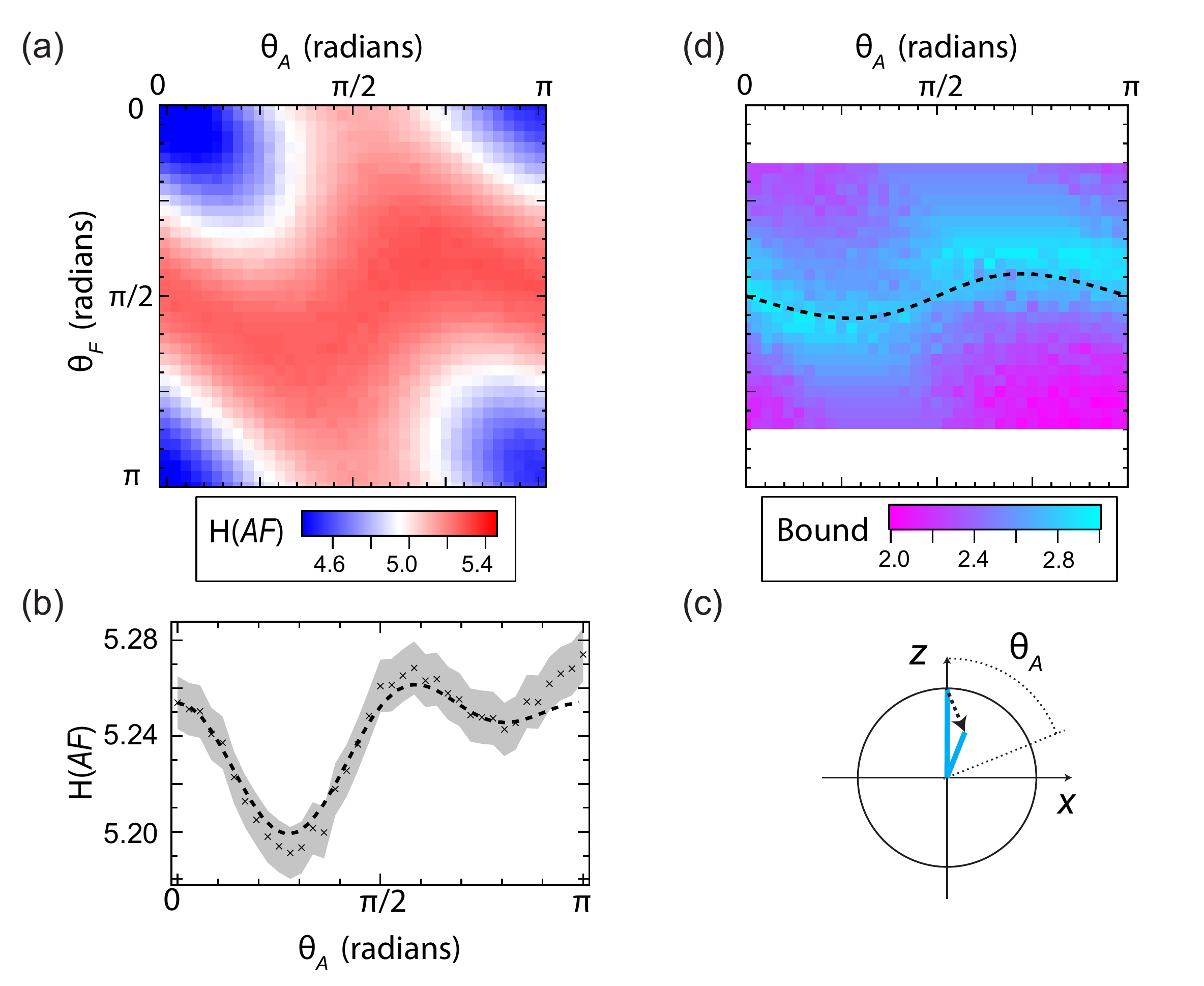}
		\caption{\jonathan{Measurements of the entropic uncertainty relation: (a) The entropy $H(AF)_\rho$. 
			(b) Detail of  $H(AF)_\rho$ versus $\theta_A$ (markers), compared to theory (dashed line), at $\theta_F = \pi / 2$. Bands indicate statistical error that results from finite sampling (approximately 140,000 repetitions per angle).		
			(c) Bloch-plane sketch indicating the $A$ measurement's backaction (dashed arrow) on the initial state.
			(d) The bound of Ineq. (\ref{EUR}).	The dashed line indicates the bound's theoretical maximum. 
		}}
		\label{fig3}
	\end{figure}
	
\emph{Results.}---Figure \ref{fig3} displays results of measuring both sides of the entropic uncertainty relation \eqref{EUR}. As above, we set $\II = \sigma_z$ and $\rho = \ketbra{0}{0}$, to achieve the tightest bound. 
Since the $\II$-measurement axis coincides with $\rho$ on the Bloch sphere, only the measurement infidelity causes the entropy $H(\mathcal{I})_\rho$ (Fig. \ref{fig2}) to contribute to the LHS of \eqref{EUR}.
We first focus on $H(AF)_\rho$, measured as a function of $\theta_F$ and $\theta_A$. The choice $\II=\sigma_z$ introduces an azimuthal symmetry that allows us to neglect rotations out of the $x$--$z$ plane.

We have already detailed the $\theta_F$-dependence of $H(AF)_\rho$ for $\theta_A=\pi/4$: Figure~\ref{fig2} showed how the weak measurement can reconcile incompatible operators. 
Here, we focus on the $\theta_A$ dependence of $H(AF)_\rho$ [Fig.~\ref{fig3}(b)].
Four effects compete to extremize $H(AF)_\rho$ as a function of $\theta_A \in [0, \pi]$, when $\theta_F= \pi / 2$.
%1: rho away from A (minimum at 0, pi; maximum at pi/2)
First, as $\theta_A$ grows from 0, the initial state’s overlap with an $A$ eigenstate decreases. $A$-measurement outcomes are sampled from an increasingly uniform distribution. This effect helps maximize $H(AF)_\rho$ at $\theta_A = \pi/2$.
%2: Back-action biases output; minimum at pi/2.
Second, as $A$ approaches $F$, the $A$ measurement's backaction biases the $F$-measurement outcome. This effect would decrease $H(AF)_\rho$ to a minimum at $\theta_A = \pi/2$, in the absence of the other effects.
%3: Inefficient detection
Third, the weak measurement partially projects the state onto the $A$ axis, dephasing the state with respect to the $A$ eigenbasis. Detection inefficiency enhances the dephasing~\cite{Suppl_Mat} and shrinks the Bloch vector [Fig.~\ref{fig3}(c)].
The $F$-measurement outcome becomes maximally biased,
minimizing $H(AF)_\rho$, when $\theta_A = \pi/4, \: 3\pi/4$. 
%4: T1 decay: minimum at 0
Fourth, readout infidelity (due to energy leakage from the qubit) raises $H(AF)_\rho$ as $\theta_A$ increases. Hence $H(AF)_\rho$ is asymmetric about $\theta_A = \pi/2$. 
% altogether: dephasing wins, a at pi/4, 3pi/4
Overall, the maxima and minima of $H(AF)$ follow from the competition between the uncertainties in the $F$-measurement and $A$-measurement outcomes. Our experimental apparatus's finite measurement efficiency~\cite{Suppl_Mat} masks the $A$ measurement’s contribution, resulting in minima at $\theta_A = \pi/4$ and $3\pi/4$.

Figure \ref{fig3}(d) displays measured values of the entropic uncertainty relation's RHS. We measure $p_{f | i}$, $p_j$, and $A_\mathrm{wv}$ in separate sets of experiments. We calculate $p_{f | i} = | \brakett{f}{i} |^2$ by preparing an $\II$ eigenstate $\ket{i}$ and measuring $F$ in each of many trials. From the frequency with which $f$ occurs, we infer the conditional probability.	
 The $p_j$ and $g_j$ in (\ref{eq_Kraus2}) are obtained from the weak-measurement calibration \cite{Suppl_Mat}. Finally, we measure the weak value $A_\mathrm{wv}$ by preparing an $\II$ eigenstate $\ket{i}$, measuring $A$ weakly, and then measuring $F$ projectively, in each of many trials. Then, we postselect on the final-measurement outcome $f$. An average of the weak-measurement outcomes $j$ is proportional to $A_\mathrm{wv}$ \cite{Suppl_Mat}.
We measure the uncertainty relation's RHS only where 
$\theta_F\in[\pi/6, \: 5 \pi/6]$, due to low postselection-success rates closer to 0 and to $\pi$.
Having measured $p_{f | i}$, $p_j$, and $A_\mathrm{wv}$ for each choice of $(i,j,f)$, we calculate the argument of the minimum in Ineq.\ \eqref{EUR}. We then identify the minimizing triple.

In Fig. \ref{fig3}(d),
the maximum of the bound, the inequality's RHS, varies sinusoidally with $\theta_A$. 
Though $F$ disagrees most with $\mathcal{I}$ at $\theta_F=\pi/2$, 
the weak $A$ measurement shifts the maximum's location.
 For example, when $\theta_A = \pi/4$, the maximally disagreeing $AF$ measurement has $\theta_F=0.53\pi$, when the measurement strength is $\delta t/\tau = 0.17$.
 When $\theta_F=\pi/2$, setting $\theta_A$ to $\pi/4$ reconciles disagreeing operators, $\sigma_z$ and $\sigma_x$.

\nicole{The weak value $A_\wv$ [Eq.~\eqref{eq_Weak_Val_Def}] 
underlies the reconciliation:
The weak-value term in Ineq.~\eqref{EUR} tends to assume negative values, lowering the bound.
Additionally, $A_\wv$ can grow anomalous, straying outside the $A$ spectrum.
However, large-magnitude $A_\wv$ values can violate the Taylor approximation that led to Ineq.~\eqref{EUR}~\cite{Halpern2018}.
As we focus on the uncertainty relation, anomalous weak values lie outside the scope of this study. 
}

Finally, we examine the bound's tightness, 
the difference between the LHS and RHS. 
The bound tightens maximally
not just at one measurement orientation,
but throughout a set of orientations near $\theta_F = \pi / 2$.
Here, the tightness is $2.45 \pm 0.05$ bits. 
\jonathan{The tightness is ideally 0.7 bits, but inefficient detection raises the entropy sum's empirical value by 1.66 bits.}
% Of the 2.45 bits, 1.66 bits arise from inefficient detection.
	
\emph{Discussion.---}We have experimentally measured an entropic uncertainty relation for strong and weak measurements~\cite{Halpern2018}, using a circuit-QED platform. A weak measurement, we have shown, can reconcile incompatible operations: up to a normalization floor, the weak measurement decreases the entropy sum on the inequality's LHS and the uncertainty bound on the RHS. This work opens operator reconciliation to feedback-free control by weak measurements, which have recently been used to control steering~\cite{Roy_19_Measurement} and pure-state preparation~\cite{Muhonen_18_Coherent} without feedback.
This work also suggests benefits of using weak measurements in applications of entropic uncertainty relations, as to quantum cryptography \cite{Coles2014}.

Mathematically, a weak value lowers the uncertainty relation's RHS. The weak value's influence is visible also in the sinusoidal variation of the RHS with the weak-measurement angle. This work therefore demonstrates a new physical interpretation of the weak value: the weak value controls the uncertainty bound on operations formed from strong and weak measurements. % Unlike some other interpretations, this one 
Whereas other interpretations have excited controversy, this interpretation is, we believe,
mathematically clear and experimentally supported.

Entropic uncertainty relations have been measured with various platforms, 
including neutrons, optics, and nitrogen-vacancy centers~\cite{Li2011,Prevedel2011,Xing2017,Demirel2019a}.
The measurements in~\cite{Demirel2019a}, though nonprojective,
are probabilistic projections.
In contrast, our measurements are weak and experimentally demonstrate 
the weak value's role in reconciling incompatible operations.
This role has only been mentioned theoretically~\cite{Halpern2018}, 
neither detailed nor experimentally tested, until now.
Uncertainty relations occupy two categories~\cite{Coles2017}, 
one centered on measurement outcomes' unpredictability~\cite{Xing2017, Demirel2019a} 
and one centered on measurements' disturbance of quantum states~\cite{Li2011, Prevedel2011}.
Our uncertainty relation occupies both categories, in the spirit of~\cite{Rozema_12_Violation}:
on the one hand, we prepare an $\II$ eigenstate $\ket{i}$ and perform the composite $AF$ measurement.
On the other hand, we take advantage of the weak $A$ measurement's disturbance of $\ket{i}$.
This work identifies weak measurements as a means of unifying the classes of uncertainty relations.

The measured uncertainty relation follows from 
simplifying an entropic uncertainty relation
for quantum information scrambling~\cite{Halpern2018}.
Quantum information \emph{scrambles} by spreading through many-body entanglement, during a nonclassical stage of equilibration~\cite{Shenker_Stanford_14_BHs_and_butterfly,Roberts_Stanford_15_Diagnosing,Maldacena_15_Bound,Kitaev_15_Simple}.
The entropic uncertainty relation for quantum-information scrambling
occupies a recent line of 
theoretical applications of weak measurements to scrambling~\cite{NYH_17_Jarzynski,NYH_18_Quasiprobability,Swingle_18_Resilience,GonzalezAlonso_19_Out,Dressel_18_Strengthening,Halpern2018,Mohseninia_19_Optimizing,DRMAS_19_Quantum, Arvidsson-Shukur2020}.
Our experiment is the first to arise from this theory.
It paves the way for characterizations of scrambling 
with weak measurements of many-body quantum systems.

\begin{acknowledgements}
\emph{Acknowledgments}---KWM acknowledges support from NSF  No. PHY- 1752844 (CAREER) and use of facilities at the Institute of Materials Science and Engineering at Washington University. NYH is grateful for funding from the Institute for Quantum Information and Matter, an NSF Physics Frontiers Center (NSF Grant PHY-1125565) with support of the Gordon and Betty Moore Foundation (GBMF-2644), and for an NSF grant for the Institute for Theoretical Atomic, Molecular, and Optical Physics at Harvard University and the Smithsonian Astrophysical Observatory. This project began at the KITP's 2018 ``Quantum Thermodynamics" conference and so was supported in part by the National Science Foundation under Grant No. NSF PHY-1748958.
\end{acknowledgements}


%%%%%% ARXIV ONLY: append supplemental  %%%%%% 
%%%%%%%%%%%%%%%%%%%%%%%%%%%%%%%%% 
	\pagebreak
	
	\newpage  
	
	\newpage
	\widetext
	\begin{center}
		\textbf{\large Supplemental Material}
	\end{center}
	%%%%%%%%%% Prefix a "S" to all equations, figures, tables and reset the counter %%%%%%%%%%
	\setcounter{equation}{0}
	\makeatletter
	\renewcommand{\theequation}{S\arabic{equation}}
	\renewcommand{\thefigure}{S\arabic{figure}}
	\renewcommand{\bibnumfmt}[1]{[#1]}
	\renewcommand{\citenumfont}[1]{#1}
	
	\section{Experimental setup}
	
	The experimental setup consists of a superconducting transmon circuit embedded in a three-dimensional copper cavity. The cavity is shielded in a copper enclosure and encased in aluminum and Cryoperm magnetic shields. The cavity is coupled to a 50 $\Omega$ transmission line that is directed, via a microwave circulator, to a Josephson parametric amplifier. The amplifier operates with $\sim 10$ MHz of instantaneous bandwidth in phase-sensitive mode.  The microwave setup is similar to that in Ref.~\cite{Murch2013}. The transmon and amplifier were fabricated with direct-write photolithography and double-angle evaporation of aluminum on an intrinsic silicon wafer.
	
	\section{Calibration measurements}
	
	To calibrate the measurement strength, we probe the cavity, near resonance, with a microwave drive. We examine the ensemble dephasing rate and the ac Stark shift, which are related to system parameters as $\Gamma_\mathrm{m} = 8\chi^2 \nbar/\kappa$ and $2\chi \nbar$, respectively. The cavity linewidth, measured in transmission through the cavity with a vector network analyzer, is $\kappa/(2\pi) = 4.5$ MHz. From these values, we infer the dispersive coupling rate, $\chi/(2\pi) = -1.5$ MHz, and the mean intracavity photon number,  $\bar{n} = 0.5$.
	
	\nicole{The coupling strength $g_j$ is real in our experiment. As a result, only the weak value's real part contributes to the uncertainty relation~\eqref{EUR}, through $\Re (g_j A_\weak)$. The weak value's imaginary part contributes only if $g_j$ is imaginary. We can set $g_j$ to be imaginary, by measuring the probe in a different basis. However, if $g_j$ and $A_\weak$ were imaginary, the $i$ in $g_j$ would cancel with the $i$ in $A_\weak$. The uncertainty bound would be the same as if $g_j$ and $A_\weak$ were real. Hence any imaginary part of $A_\weak$ contributes no new physics to the uncertainty relation. We therefore study real $g_j$ and $A_\weak$ without loss of generality.
	}
	
	The quantum efficiency reduces the signal-to-noise ratio with which we resolve the quantum states. We treat the quantum efficiency as noise added to our measurement signal. To calibrate the quantum efficiency, 
	we prepare the qubit in the state $\rho_0 = \ketbra{0}{0}$,
	then measure $A = \sigma_z$ weakly, in each of many trials.
	In another set of trials, $\rho = \ketbra{1}{1}$. The probability of obtaining outcome $j$ follows from introducing the quantum efficiency $\eta$ into the Kraus operator~\eqref{eq_Kraus1}:
	$$
	p_{j|\rho} 
	= \mathrm{Tr} (\rho K_j^\dagger K_j) 
	= \left(  \frac{\delta t \, \eta}{{2 \pi \tau}}  \right)^{1/2} 
	\exp \left(  -\frac{\delta t \, \eta}{2\tau} [j \pm 1]^2  \right),
	$$
	with $\rho_0$ corresponding to $\pm = +$ and $\rho_1$ corresponding to $\pm = -$. We have introduced the ensemble characteristic measurement time $\tau = 1/(2\Gamma_m) =  \kappa/(16 \chi^2 \nbar )$.
	
	In the experiment, we measure only $\sigma_z$ directly. To measure different observables, we rotate the state before and after the $\sigma_z$ measurement.

	\section{Minima of $H(AF)$}
	In the main text, we discussed the extrema of $H(AF)_\rho$ as functions of $\theta_A \in [0, \pi]$, with $\theta_F=\pi/2$ and $\rho = \ketbra{0}{0}$.	We observed that ensemble-level dephasing, induced by the weak $A$ measurement, biases the $F$-measurement outcome. As a result, $H(AF)_\rho$ minimizes at $\theta_A = \pi/4$ and $3\pi/4$. Surprisingly, a direct calculation of $H(AF)_\rho$ exhibits \emph{maxima} at $\theta_A = \pi/4$ and $3\pi/4$.
		The seeming contradiction arises because the changes in the joint distribution over $(j, f)$ outcomes are small. Inefficient detection masks the small changes.
	 Hence ensemble-level evolution dominates the qualitative behavior of $H(AF)_\rho$.
	
	Our measurements agree fairly well with a model that includes inefficient detection. We model inefficient detection via a two-measurement process \cite{Jacobs2006}. One measurement record describes the detected outcome, and the other measurement record describes an outcome lost to the environment. 
Averaging over all possible values of the lost outcome increases the detected outcome's variance.

	\section{How the weak value is measured}
	
	Our measurement strategy relies on a proportionality proved in the next section: The real part of the weak value, $\Re (A_\mathrm{wv})$, is proportional to the average preselected and postselected detector outcome, $\langle \langle j \rangle \rangle$. To measure $\langle \langle j \rangle \rangle$, we prepare a state $\rho_0$, rotate it downward toward the $x$-axis through an angle $-\theta_A$, measure $\sigma_z$ weakly for 250 ns, and rotate the state back. This measurement process yields an outcome $j$. Finally, we measure the observable $F$ projectively, obtaining an outcome $f$. We repeat this protocol in many trials. Processing the outcome statistics, as dictated in the next section, yields $\langle \langle j \rangle \rangle$.

	\section{Proportionality between the weak value $A_\wv$ 
		and an average of $j$}
	\label{sec_Weak_Val_Prop}
	
	Consider preparing a system of interest in a state $\ket{i}$, 
	coupling the detector to a system observable 
	$A = \sum_a a \ketbra{a}{a}$,
	measuring the detector strongly,
	and then measuring a system-of-interest observable
	$F =  \sum_f f \ketbra{f}{f}$ strongly.
	Let $f$ denote the $F$ measurement's outcome.
	The measurement of the detector yields a random variable $j$.
	An average of $j$ is proportional to the weak value $A_\wv$.
	We define the average and derive the proportionality constant here.
	
	We choose for our model to have two features that merit explaining.
	First, we assume that the system of interest is a qubit, as in the main text.
	$A$ represents a Pauli operator, which squares to the identity operator:
	$A^2 = I$.
	Second, $j$ is continuous ideally, and we model it as continuous, 
	here, for convenience.
	In an experiment, however, $j$ values are binned,
	so the variable is discretized.
	
	For ease of reading, we change notation from  
	$p_{j, f}$ and $p_{f | i}$ to $p(j, f)$, $p(f | i)$.
	The probability of obtaining outcome $f$, conditioned on 
	one's having prepared $i$ and obtained $j$, equals 
	% < /f >
	\begin{align}
	p (f | i, j)
	% % %
	& =  | \bra{f} K_j \ket{i} |^2  \\
	% % %
	\label{eq_prf_1}
	& =   p_j   \left\{  p(f | i)
	+ 2 \Re ( g_j  \brakett{i}{f}  \bra{f} A \ket{i} )
	+ O \left(  \left[ g_j  \right]^2  \right)  \right\} .
	\end{align}
	We have substituted in for $K_j$ from Eq.~\eqref{eq_Kraus2}.
	$p(f | i) = | \brakett{f}{i} |^2$ denotes the conditional probability that
	preparing $\ket{i}$ and measuring $F$ yields $f$.
	The $\bra{f} A \ket{i}$ equals the numerator in the definition of $A_\wv$  
	[Eq.~\eqref{eq_Weak_Val_Def}].
	We substitute $\brakett{f}{i} A_\wv$ into Eq.~\eqref{eq_prf_1}.
	Dividing each side of the equation by $p(f | i)$ simplifies the RHS:
	\begin{align}
	\frac{ p(f | i, j) }{ p(f | i) }
	% % %
	\label{eq_prf_3}
	=   p_j  \left\{  1  +
	2  \Re \left( g_j  A_\wv  \right)   
	+  O  \left(  \left[  g_j  \right]^2  \right)  \right\} .
	\end{align}
	
	Bayes' theorem offers an interpretation of the LHS of Eq.~\eqref{eq_prf_3}.
	We derive the interpretation by writing two expressions for
	the joint probability $p(i, j, f)$ of preparing $\ket{i}$, then obtaining $j$ and $f$,
	\begin{align}
	& p(i, j, f)  =  p(f | i, j)  \,  p(i, j) 
	=  p(f | i, j)  \,  p(j | i)  \,  p(i) 
	\nonumber \\ & \qquad \quad
	=  p(f | i, j)  \,  p(j | i) 
	\label{eq_Joint_P1}
	% % %
	\quad \text{and} \quad \\
	% % %
	& p(i, j, f)  =  p(j | i, f)  \,  p(i, f)
	=  p(j | i, f)  \,  p(f | i)  \,  p(i)
	\nonumber \\ & \qquad \quad
	\label{eq_Joint_P2}
	=  p(j | i, f)  \,  p(f | i) .
	\end{align}
	$p(i) = 1$ because our protocol requires a deterministic preparation of $\ket{i}$.
	We equate~\eqref{eq_Joint_P1} and~\eqref{eq_Joint_P2},
	then solve for the LHS of Eq.~\eqref{eq_prf_3}:
	\begin{align}
	\frac{ p(f | i, j) }{ p(f | i) }
	=  \frac{ p(j | i, f) }{ p(j | i) }  \, .
	\end{align}
	This ratio reflects the impact of the $f$---of the postselection---on the probability.
	
	We simplify the calculation of the RHS of Eq.~\eqref{eq_prf_3},
	by stipulating that the detector be calibrated as follows.
	Suppose that the detector is prepared,
	is not coupled to the system of interest,
	and is measured strongly.
	The average outcome $j$ is set to zero:
	$\int dj \cdot j \cdot p_j = 0$.
	This calibration condition amounts to a choice of a plot's origin.
	We invoke this condition upon integrating $j$ against 
	the RHS of Eq.~\eqref{eq_prf_3}:
	\begin{align}
	\expval{ \expval{j} }
	% % %
	& := \int_{-\infty}^\infty dj  \cdot  j  \:  \frac{ p(j | i, f) }{ p(j | i) }  \\
	% % %
	\label{eq_j_1}
	& = \int_{-\infty}^\infty  dj  \cdot  j  \, p_j   \:
	\left\{  2  \Re ( g_j  A_\wv )
	+ O \left(  \left[  g_j  \right]^2  \right)   \right\}.
	\end{align}
	
	We simplify and decompose the integral
	by specializing to our weak-measurement setup.
	The outcome-dependent coupling $g_j$ is real, so
	$\Re ( g_j  A_\wv )  =  g_j  \Re ( A_\wv )$.
	We calculate $p_j$ and $g_j$ by 
	Taylor-approximating the RHS of Eq.~\eqref{eq_Kraus1} 
	and comparing the result with Eq.~\eqref{eq_Kraus2}:
	\begin{align}
	\label{eq_p_j_Form}
	& \sqrt{ p_j }  
	=  \left( \frac{ \delta t }{ 2 \pi \tau } \right)^{1/4}  
	\exp \left( - \frac{ \delta t }{ 4 \tau }   \left[ j^2 + 1 \right]  \right) ,
	% % %
	\quad \text{and} \\
	% % %
	\label{eq_g_j_Form}
	& g_j  =  \frac{ \delta t }{ 2 \tau }  \, j   .
	\end{align}
	We substitute from these equations into Eq.~\eqref{eq_j_1} 
	and evaluate the integral.
	
	In our experiment, $g_j \leq 0.3$, and $ | A_\wv | \approx 1 $. 
	Hence the $O \left(  \left[  g_j  \right]^2  \right)$ term in Eq.~\eqref{eq_j_1}
	is about an order of magnitude less than the first term, and
	\begin{align}
	\boxed{
		\expval{ \expval{j} }
		\approx  \Re (  A_\wv  )  \,
		e^{ - \delta t / (2 \tau) }   }  \, .
	\end{align}
	Since the coupling is weak, $\frac{ \delta t }{ \tau }  \ll  1$,
	$e^{ - \delta t / (2 \tau) }  \approx 1$, and
	$\expval{ \expval{j} }
	\approx  \Re (  A_\wv  )$.

\nicole{	
\section{Derivation of the entropic uncertainty relation for weak and strong measurements}

We will recapitulate the derivation of Ineq.~\eqref{EUR},
which first appeared on p. 7 of~\cite{Halpern2018}.
Following~\cite{Halpern2018}, we present a formalism
more general than that of the main text's qubit.
Let $\II  =  \sum_i \lambda_i  \Pi^\II_i$, 
$\A  =  \sum_a a \Pi^\A_a$, and 
$\F  =  \sum_f f  \Pi^f_f$ 
be eigendecompositions of observables of a quantum system.

Let us define two POVMs, I and II.
POVM I is a composite measurement:
$\A$ is measured weakly, and then $\F$ is measured projectively.
Suppose that the weak measurement yields an outcome $j$
and the $\F$ measurement yields an outcome $f$.
The Kraus operator
\begin{align}
   \label{eq_Kraus_I}
   \sqrt{ M^\povmI_{j, f} }
   :=  \Pi^F_f  K_j 
\end{align}
evolves the system-of-interest state.
The weak-measurement Kraus operator $K_j$ 
is defined in Eq.~\eqref{eq_Kraus2}.
POVM II consists of a strong measurement of $\II$.
If outcome $\lambda_i$ obtains,
the measurement evolves the system-of-interest state 
with the Kraus operator
\begin{align}
   \label{eq_Kraus_II}
   \sqrt{ M^\povmII_i }
   :=  \Pi^\II_i .
\end{align}

More notation is in order.
POVM I yields outcome $(j, f)$ with probability
$q^\povmI_{j, f}
   = \Tr \left(  \sqrt{ M^\povmI_{j, f} }  \:
   \rho  
   \sqrt{ M^\povmI_{j, f} }^\dag  \right) .$
The corresponding probability distribution has a Shannon entropy
$H (AF)_\rho
   = - \sum_{j, f}  q^\povmI_{j, f}  \log q^\povmI_{j, f} .$
POVM II yields outcome $\lambda_i$ with a probability
$q^\povmII_i
   = \Tr  \left(  \sqrt{ M^\povmII_i }  \:
   \rho
   \sqrt{ M^\povmII_i }^\dag  \right) ,$
which corresponds to a Shannon entropy
$H( \II )_\rho
   = - \sum_i  q^\povmII_i  \log q^\povmII_i .$
The weak value is defined as
\begin{align}
   \label{eq_Weak_Val_Def_app}
   A_\weak
   := \frac{ \Tr \left( \Pi^F_f A \Pi^\II_i \right) }{
                \Tr \left( \Pi^F_f \Pi^\II_i  \right)
                \Tr \left( \Pi^\II_i  \right) }  \, .
\end{align}

A generalized entropic uncertainty relation for POVMs
is proved in~\cite{Tomamichel_12_Framework,Krishna_01_Entropic}.
We substitute in POVMs I and II:
\begin{align}
   \label{eq_Op_Norm_Help0}
   H (AF)_\rho  +  H( \II )_\rho
   \geq - \log c \left( \Set{ M^\povmI_{j, f} },  \Set{ M^\povmII_i }  \right) .
\end{align}
The maximum overlap is defined as
\begin{align}
   \label{eq_Max_Over_Supp}
   c \left( \Set{ M^\povmI_{j, f} },  \Set{ M^\povmII_i }  \right)
   % % %
   := \max_{j, f, i}  \Set{  \left\lvert \left\lvert
   \sqrt{  M^\povmI_{j, f}  }  \:
   \sqrt{  M^\povmII_i  }
   \right\rvert  \right\rvert^2   }  .
\end{align}
The operator norm is defined as
\begin{align}
   \left\lvert \left\lvert
   \sqrt{  M^\povmI_{j, f}  }  \:
   \sqrt{  M^\povmII_i  }
   \right\rvert  \right\rvert
   % % %
   \label{eq_Op_Norm}
   :=  \lim_{\alpha \to \infty}  \Set{  \Tr  \left(  \sqrt{
   \left[ \sqrt{ M^\povmI_{j, f} }   \:  \sqrt{ M^\povmII_i }  \right]^\dag
   \left[  \sqrt{ M^\povmI_{j, f} }   \:  \sqrt{ M^\povmII_i }  \right]
   }^\alpha  \right)  }^{1 / \alpha } .
\end{align} 
The outer square-root equals
\begin{align}
   \sqrt{ \sqrt{ M^\povmII_i  }^\dag
            \sqrt{ M^\povmI_{j, f}  }^\dag
            \sqrt{ M^\povmI_{j, f} }
            \sqrt{ M^\povmII_i  } }
   % % %
   = \sqrt{ \Pi^\II_i  \left( K_j \right)^\dag
   \Pi^F_f   \Pi^F_f   K_j  \Pi^\II_i  }
   % % %
   \equiv  \sqrt{\Op} .
\end{align}
The operator $\Op$ is Hermitian
and so has an eigendecomposition.

Much of the text from here until just after Eq.~\eqref{eq_Op_Norm_Help3b}
consists of quotations from 
Supplementary~Note~1 of~\cite{Halpern2018}.
We omit quotation marks for ease of reading.
The eigenvalues of $\Op$ are real and nonnegative, for two reasons.
First, the eigenvalues are the squares of the singular values of
$\sqrt{  M^\povmII_{j, f}  }    \sqrt{  M^\povmII_i  }$.
All singular values are real and nonnegative.
Second, $\Op$ is proportional to a quantum state,
constructed as follows.
$\Pi^\II_i / \Tr ( \Pi^\II_i )$ represents the state that is
maximally mixed over the eigenvalue-$\lambda_i$ 
eigenspace of $\II$.
Imagine subjecting this state to the quantum channel
defined by the operation elements 
$\Set{ K_j }$, then measuring $\F$ projectively.
The resultant state, $\sigma_f$, is proportional to $\Op$.
The proportionality factor equals $1 / \Tr ( \Op )$.
The trace equals the joint probability that 
(i) $j$ labels this realization of the channel's action
and (ii) the $\F$ measurement yields outcome $f$.
Since $\sigma_f = \Op / \Tr (\Op)$ is positive-semidefinite
and $\Tr(\Op)$ equals a probability
(which is real and nonnegative),
the eigenvalues of $\Op$ are real and nonnegative.

The eigenvectors of $\Op$ are eigenvectors of $\Pi^\II_i$.
$\Pi^\II_i$ has two distinct eigenvalues $\lambda_i$:
0, of degeneracy $\Tr ( \id - \Pi^\II_i )$, 
and 1, of degeneracy $\Tr ( \Pi^\II_i )$.
Let $\Lambda_{i}^r$ denote the $r^\th$ $\Op$ eigenvalue
associated with any eigenvector in 
the $\lambda_i$ eigenspace of $\Pi^\II_i$.
If $d_i$ denotes the degeneracy of $\Lambda^r_i$,
then $r = 1, 2, \ldots, d_i$.
(We have omitted the $i$-dependence 
from the symbol $r$ for notational simplicity.)
Every eigenvalue-0 eigenvector of $\Pi^\II_i$ is
an eigenvalue-0 eigenvector of $\Op$:
$\Lambda_0^r = 0 \;  \;  \;
\forall r = 1, 2, \ldots, \Tr ( \id - \Pi^\II_i )$.
Hence $\Op$ eigendecomposes as
\begin{align}
   \Op = \sum_{\lambda_i = 0}^1  
             \sum_{r = 1}^{d_i}
             \Lambda_{i}^r  \Pi^r_{i}
   = 0 \left( \id - \Pi^\II_i  \right)
      + \sum_{r = 1}^{d_i}  
         \lambda_1^r  \Pi^r_1 .
\end{align}
The sum over $\lambda_i$ is equivalent to a sum over $i$.

We use this eigenvalue decomposition to evaluate 
the RHS of Eq.~\eqref{eq_Op_Norm},
working from inside to outside.
The outer square-root has the form
$\sqrt{\Op} = \sum_{\lambda_i = 0}^1  
\sum_{r = 1}^{d_i}
\sqrt{ \Lambda_1^r }  \:  \Pi_{\lambda_i}^r$.
The projectors project onto orthogonal subspaces, so
$\left( \sqrt{\Op} \right)^\alpha
= \sum_{r = 1}^{d_i}  
\left( \Lambda_1^r \right)^{\alpha / 2}  \Pi_1^r$.
We take the trace, 
$\Tr \left( \left[ \sqrt{\Op} \right]^\alpha \right)
= \sum_{r = 1}^{d_i}  
\left( \Lambda_1^r  \right)^{\alpha / 2}$,
then exponentiate:
$\Set{  \Tr \left(  \left[  \sqrt{\Op}  \right]^\alpha  \right)  }^{1 / \alpha}
=  \left[  \sum_{r = 1}^{d_i}
\left( \Lambda_1^r  \right)^{\alpha / 2}  
\right]^{1 / \alpha}$.
The limit as $\alpha \to \infty$ gives the RHS of Eq.~\eqref{eq_Op_Norm}:
\begin{align}
   \left\lvert \left\lvert
   \sqrt{  M^\povmI_{j, f}  }  \:
   \sqrt{  M^\povmII_i  }
   \right\rvert  \right\rvert
   % % %
   \label{eq_Op_Norm_Help1}
   & =  \lim_{\alpha \to \infty} 
   \Set{ \Tr \left(  \left[  \sqrt{\Op}  \right]^\alpha  \right)  }^{1 / \alpha} \\
   % % %
   \label{eq_Op_Norm_Help2}
   & = \lim_{\alpha \to \infty}
   \left[  \sum_{r = 1}^{d_i}  
   \left( \Lambda_1^r  \right)^{\alpha / 2}  \right]^{1 / \alpha} .
\end{align}

Only the greatest eigenvalue survives:
$\left\lvert \left\lvert
   \sqrt{  M^\povmI_{j, f}  }  \:
   \sqrt{  M^\povmII_i  }
   \right\rvert  \right\rvert
   =  \sqrt{ \Lambda_1^\Max }$.
But $\Lambda_1^\Max$ is neither a parameter
chosen by the experimentalist
nor obviously experimentally measurable.
Hence bounding the entropies 
with $\Lambda_1^\Max$ appears useless.

Probabilities and weak values are measurable,
and $\Tr (\Op)$ equals a combination of them.
We therefore seek to shift the $\Tr$ of Eq.~\eqref{eq_Op_Norm_Help1}
inside the $[.]^\alpha$ and the $\sqrt{.} \; .$
Equivalently, we seek to shift the $\sum$ of Eq.~\eqref{eq_Op_Norm_Help2}
inside the $( . )^{\alpha / 2}$.
We do so at the cost of introducing an inequality:
\begin{align}
   \label{eq_Op_Norm_Help3}
   \sum_{r = 1}^{d_i}  ( \Lambda_1^r )^{\alpha / 2}
   \leq  \left(  \sum_{r = 1}^{d_i}  \Lambda_1^r  \right)^{\alpha / 2}
\end{align}
for all $\alpha / 2 \geq 1$.
This inequality follows from the Schatten $p$-norm's monotonicity.
The Schatten $p$-norm of an operator is defined as
$|| \sigma ||_p  :=  
\left[ \Tr \left( \sqrt{ \sigma^\dag \sigma }^{\; p}  \right)  \right]^{1 / p}$,
for $p \in [1, \infty)$.
As $p$ increases, the Schatten norm decreases monotonically:
\begin{align}
   \label{eq_Schatten_Mon}
   || \sigma ||_p
   \leq || \sigma ||_q
   \quad \text{if} \quad
   p \geq q .
\end{align}
Let $p = \alpha / 2$ and $q = 1$.
Raising each side of Ineq.~\eqref{eq_Schatten_Mon}
to the $\alpha / 2$ power yields Ineq.~\eqref{eq_Op_Norm_Help3}.
Applying Ineq.~\eqref{eq_Op_Norm_Help3} to Eq.~\eqref{eq_Op_Norm_Help2}
bounds the operator norm as
\begin{align}
   \label{eq_Op_Norm_Help3b}
   \left\lvert \left\lvert
   \sqrt{  M^\povmI_{j, f}  }  \:
   \sqrt{  M^\povmII_i  }
   \right\rvert  \right\rvert
   % % %
   \leq \sqrt{ \Tr \left(  
   \Pi^\II_i  \left[ K_j \right]^\dag
   \Pi^\F_f   K_j   \right)  } \, .
\end{align}
We have invoked the trace's cyclicality
and the $\F$ projector's equality to its own square:
$\left( \Pi^F_f  \right)^2  =  \Pi^\F_f$.

Combining Ineq.~\eqref{eq_Op_Norm_Help3b} with Eq.~\eqref{eq_Max_Over_Supp} yields
\begin{align}
   c \left( \Set{ M^\povmI_{j, f} },  \Set{ M^\povmII_i }  \right)
   % % %
   & \leq  \max_{j, f, i}  \Set{  \Tr \left(
   \Pi^\II_i  [K_j]^\dag  \Pi^\F_f  K_j  \right)  } .
\end{align}
We substitute in for $K_j$ from Eq.~\eqref{eq_Kraus2},
multiply out, and invoke the trace's linearity:
\begin{align}
   c \left( \Set{ M^\povmI_{j, f} },  \Set{ M^\povmII_i }  \right)
   % % %
   & \leq \max_{j, f, i}  \Set{ 
   p_j  \Tr \left(  \Pi^\II_i  \Pi^\F_f  \right)
   + 2 p_j \:  \Tr \left(  \Pi^\F_f  \Pi^\II_i  \right)
   \Tr \left( \Pi^\II_i  \right)
   \Re \left( g_j  A_\weak  \right)
   + O \left(  p_j  \:  g_j  \right)^2   } .
\end{align}
We take the logarithm of each side of the inequality:
\begin{align}
   \log c ( \ldots )  
   % % %
   & \leq  \log ( \max \{ \ldots \} )
   = \max \Set{ \log ( \ldots ) } ,
\end{align}
by the logarithm's monotonicity.
Negating each side yields
\begin{align}
   - \log c (\ldots) 
   % % %
   \geq - \max \Set{ \log ( \ldots ) }
   % % %
   \label{eq_Op_Norm_Help4}
   = \min \{ - \log ( \ldots ) \} .
\end{align}

Let us combine Ineqs.~\eqref{eq_Op_Norm_Help4}
and~\eqref{eq_Op_Norm_Help0}.
We factor out $p_j  \Tr ( \Pi^\II_i \Pi^F_f )
\equiv  p_{f | i}  \,  p_j$,
then apply the log law for multiplication:
\begin{align}
   \label{eq_Op_Norm_Help5}
   H (AF)_\rho  +  H(\II)_\rho
   % % %
   \geq \min_{j, f, i}  \Set{
   - \log \left( 
   p_{f | i}  \,  p_j  \right)
   - \log \left(  1  + 
   2 \Tr \left( \Pi^\II_i \right)   \:
   \Re \left( g_j  \A_\weak \right)  \right)
   + O \left(  p_j  g_j^2  \right)  } .
\end{align}
We Taylor-approximate the second logarithm,
after changing its base from 2 to $e$:
\begin{align}
   - \log \left(  1  + 
   2 \Tr \left( \Pi^\II_i \right)   \,
   \Re \left( g_j  \A_\weak \right)  \right)
   % % %
   = \frac{ -2}{\ln 2}  \:
   \Tr \left( \Pi^\II_i \right)  \,
   \Re \left( g_j  \A_\weak \right) .
\end{align}
Substituting into Ineq.~\eqref{eq_Op_Norm_Help5}
yields the main text's Ineq.~\eqref{EUR}.
}

 %\bibliography{full_bib}


\onecolumngrid
%

%	\bibliography{sm_ref}
\end{document}